# High contrast imaging of reversibly switchable fluorescent proteins via temporally unmixed Multispectral Optoacoustic Tomography (tuMSOT)

Andre C Stiel,[1]+ X. Luís Deán-Ben,[2]+ Yuanyuan Jiang,[2,3], Vasilis Ntziachristos [2,4,5], Daniel Razansky,[2,5] and Gil G. Westmeyer[2,3,5,6]*

[1]*Research Group Protein Design – Max-Planck-Institute for Developmental Biology, Tübingen, Germany*
[2]*Institute for Biological and Medical Imaging (IBMI) and* [3]*Institute of Developmental Genetics (IDG), Helmholtz Zentrum München, Neuherberg, Germany,* [4]*Chair for Biological Imaging,* [5]*School of Medicine and* [6]*Department of Nuclear Medicine, Technische Universität München (TUM), Munich, Germany*
+*these authors contributed equally to this work*
*Corresponding author: gil.westmeyer@tum.de*

*Photocontrol of reversibly switchable fluorescent proteins (RSFPs) was used to program optoacoustic signal time courses that were temporally unmixed to increase the proteins' contrast-to-noise-ratios (CNRs) in optoacoustic imaging. In this way, two variants of the RSFP Dronpa with very similar optoacoustic spectra could be readily discriminated in the presence of highly absorbing blood. Addition of temporal unmixing to multispectral optoacoustic tomography (tuMSOT) in conjunction with synthetic or genetically controlled photochromic contrast agents and customized photoswitching schedules can increase the performance of multiplexed and high contrast molecular optoacoustic imaging.*

Synthetic as well as genetically encoded fluorophores have become indispensable tools for biomedical research and diagnostics enabling a multitude of applications based on *in vitro* and *in vivo* detection and imaging schemes. Dedicated imaging instrumentation has been designed to differentiate a specific fluorescent contrast agent from background signals based on its fluorescence spectrum. This is usually achieved using optical filters that limit the bandwidth of the contrast agent's emission spectrum and/or excitation wavelengths. Sampling a larger portion of the spectrum by multiple filters or spectral detectors can effectively inform spectral unmixing algorithms to substantially increase the contrast-to-noise ratio (CNR) of fluorescent contrast agents and also routinely allow for multiplexing with several fluorophores at once. Optoacoustic (OA, or photoacoustic) imaging is an *in vivo* imaging method that exceeds the penetration limits imposed by light diffusion for high-resolution imaging of optical contrast in biological tissue. By using laser illumination at multiple optical wavelengths, multispectral optoacoustic tomography (MSOT) further enables mapping the distribution of specific absorbers in a sample via resolving their optoacoustic spectra [1]. MSOT can thus detect well-established synthetic dyes such as indocyanine green (ICG) or genetically encoded fluorescent proteins (FPs) such as green fluorescent protein (GFP) via their photoabsorbing property. However, the presence of strong endogenous absorbers, most notably blood, makes the localization of specific photoabsorbing contrast agents challenging, in particular for wavelengths below the near-infrared (NIR) range [2]. A subset of FPs and organic dyes (*e.g.* rhodamine spiroamides [3]) change their photophysical properties as a function of incident light. For the former, three different types of photoresponses can be distinguished: photoactivatable FPs (PAFPs) that can be activated from a non-fluorescent state to become fluorescent, photoswitchable FPs (PSFPs) whose fluorescence can be switched to emit at a different wavelength and reversibly switchable FPs (RSFPs) that can be switched reversibly between a fluorescent and a non-fluorescent state [4]. These properties are facilitated by permanent covalent (for PA- and PSFPs) or reversible modifications to the chromophores (for RSFPs) that result in altered fluorescence as well as absorption characteristics. Two especially well characterized RSFPs are Dronpa [5] and its fast-switching fatigue resistant variant Dronpa-M159T [6]. For both variants, green fluorescence can be excited by illumination with blue light that also switches them from their fluorescent equilibrium state to a non-fluorescent dark-state (Fig. 1A). This is accompanied by a change of the absorption spectrum featuring a hypsochromic shift of the absorption maximum. Subsequent irradiation with violet light results in reactivation of the fluorescent state, with the process being fully reversible for a number of cycles. The proteins differ in the kinetics of their off-switching; under identical illumination intensities, the variant Dronpa-M159T is several times faster as compared to the wildtype Dronpa depending on the exact experimental conditions ([6] ~1100x, [7] ~59x). The on-switching is only slightly affected by this acceleration with a two times faster rate for the variant. The variant also exhibits a far superior fatigue resistance compared to wildtype. Traditionally, photoresponsive chromo-/fluorophores have been used in tracking schemes [8] and single molecule studies [9], as well as for so-called super-resolution microscopy techniques that enable fluorescence imaging well beyond the diffraction limit [10,11,12,13]. It has furthermore been shown that the photocontrol of their fluorescence signal can be used in an optical lock-in method or Fourier domain scheme to enhance signal detection [14,15].

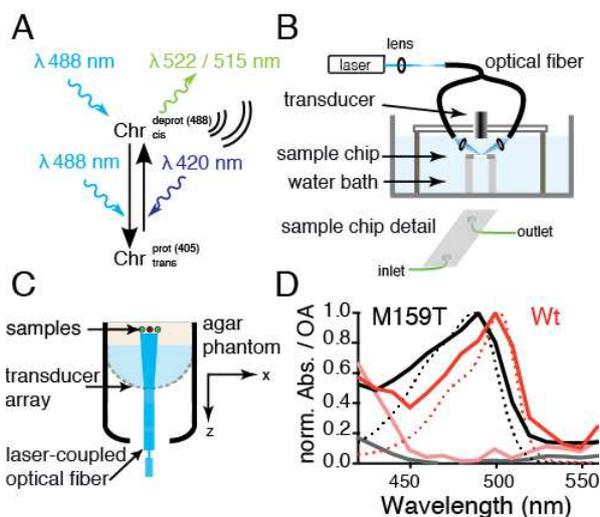

Fig 1 Photophysical properties of Dronpa and Dronpa-M159T. **(A)** Scheme depicting the switching mode of Dronpa for fluorescent and optoacoustic readout. **(B)** Cross-sectional technical drawing of the optoacoustic spectrometer (**C**) Cross-sectional technical sketch depicting the 3D optoacoustic imaging system. (**D**) OA-spectra (solid lines) of Dronpa (red) and Dronpa-M159T (black) in the equilibrium state and after prolonged illumination with 488 nm (light red and gray) as well as absorption spectra in the equilibrium state (dotted lines) are shown.

We here report the use of the photoresponsive OA signal of Dronpa and Dronpa-M159T to enhance their contrast for OA imaging. This is achieved by photoswitching the compounds according to defined schedules that elicit programmed time-varying OA signals that can be temporally unmixed to enhance localization of the specific contrast. We show that this temporally unmixed MSOT technique (tuMSOT) can differentiate well between Dronpa variants despite their highly similar optoacoustic spectra and clearly distinguish them from strong background absorbers such as blood.

To generate photocontrolled optoacoustic signal time courses, proteins of Dronpa and its fast-switching fatigue resistant variant Dronpa-M159T (purified from *E. coli*) were injected into an acoustically coupled flow chip (µ-Slide I 0.2 Luer, hydrophobic, uncoated, IBIDI, Munich, Germany) and their OA signals were detected using a single-element transducer (V382-SU, 3.5 MHz, Olympus, Hamburg, Germany). The sample was illuminated from two sides via fiberoptic light guides (Fig. 1B) coupled to a pulsed laser source delivering pulses of 10 ns duration with a pulse repetition rate of 50 Hz (SpitLight, Innolas, Krailling, Germany). A third fiber output was used to measure the laser intensity as a function of wavelength with a laser power meter (FieldMaxII-TOP, Coherent Inc. Santa Clara, USA) to correct for pulse-to-pulse energy variations. We first recorded the absorbance and OA spectra of Dronpa and Dronpa-M159T in their equilibrium states (on-state) and observed good agreement between the modalities for both proteins and an optoacoustic spectral shift between wildtype and variant of only ~10 nm (Fig. 1D). Upon switching the proteins to the off-state using prolonged irradiation at 488 nm, the bathochrome peak, corresponding to the deprotonated chromophore, shows a reduced OA signal in agreement with previous fluorescence measurements [16]. We next evaluated if defined OA signal time courses could be observed from photoswitching Dronpa with different photoswitching schedules (Fig. 2). We used 488 nm as readout/switch-off wavelength and 420 nm to switch the proteins back to their on-state (420 nm is the lowest available wavelength on the tunable Spitlight laser while the optimal wavelength for on-switching would have been 405 nm). Under the applied switching schedule Dronpa-M159T showed an approximately four-fold higher change in signal amplitude upon switching than wild type Dronpa as well as a much faster signal decay (Fig. 2A). Control over the switching behavior of the fast-switching Dronpa-M159T was further exemplified by applying different photoswitching schedules varying in the number of laser pulses at the on- and off-switching wavelength (Fig. 2B and C) as well as a more complex on-switching/readout scheme (Fig. 2D), which showcases the flexibility in programming specific time profiles in the optoacoustic signals. In a control experiment, the proteins were switched off completely with 488 nm illumination and left in the dark for approximately 30 seconds to estimate the fraction of spontaneous relaxation to the equilibrium on-state. No signal increase was detected for wildtype Dronpa, consistent with its very slow dark relaxation of $t_{1/2}$ ~840 minutes [6]. The signal for Dronpa-M159T only increased to 0.6 % of that observed after 100 pulses of 420 nm illumination, indicating that dark relaxation was negligible for the chosen photocontrol schedules. The recorded profiles of Dronpa-M159T showed no photobleaching or switching fatigue, which is in line with the published robustness of the protein [6].

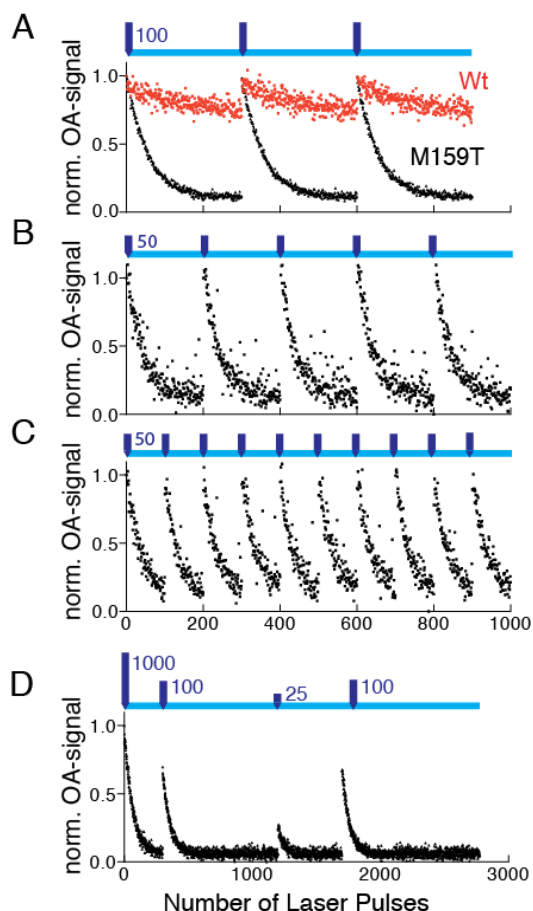

Fig. 2. Photocontrolled optoacoustic signal trajectories. **(A)** Comparison of signal time courses (60 averages) from Dronpa-Wt (red) and Dronpa-M159T (black). The photocontrol schedule is visualized by light blue bars indicating illumination with Off-switching/read-out light at 488 nm and violet bars denoting the number of laser pulses delivered at 420 nm to switch the proteins back on. The signal detected during illumination with 420 nm was cropped from the time courses. **(B/C)** Dronpa-M159T derived signal time courses (50 averages) under two periodic photocontrol schedules. **(D)** Example of a more complex switching schedule with Dronpa-M159T (20 averages). The laser pulse repetition frequency was 50 Hz for all experiments.

We were subsequently interested in determining how the photocontrolled temporal profiles could be used to effectively unmix the two Dronpa variants from each other and from blood serving as a proxy for a dominant absorber in biological tissue with spectral overlap. To this end, we imaged pure Dronpa proteins together with undiluted pig blood (~120 g/l of hemoglobin) filled in tubes immersed in a water phantom positioned in a portable three-dimensional OA imaging system (Fig. 1C) containing a transducer array with 256 ultrasonic elements [17]. As shown in the optoacoustic image in Figure 3A, blood generated an approximately 10-fold higher signal than the two Dronpa proteins (~70 µM) whose signal is only barely distinguishable from background. The corresponding Signal-to-Noise Ratios (SNRs) were calculated from the indicated regions of interest (ROIs) with regions not containing sample (top and bottom ROIs) defined as background (Fig. 3C, black and red bars). We also determined the Contrast-to-Noise-Ratio (CNR) for each Dronpa variant with respect to blood by computing the absolute value of the differences in the averaged OA signals in ROIs containing the Dronpa proteins or blood, respectively. The resulting negative CNR values for the Dronpa variants with respect to blood are due to the fact that the blood signal was an order of magnitude higher than that of the proteins. However, by utilizing the temporal information contained in the photocontrolled OA signal time courses (measured at both wavelengths) via temporal unmixing the proteins could be clearly distinguished from each other and from blood with high CNRs in the resulting unmixed images (Fig. 3C). Temporal unmixing was performed by Vertex Component Analysis (VCA) [2,18], an unsupervised linear unmixing algorithm that efficiently extracts components (endmembers) via an iterative geometrical method. To demonstrate robustness and applicability of temporal unmixing of photocontrolled OA signal trajectories, we additionally imaged whole bacterial suspensions of *E. coli* expressing fast-switching Dronpa-M159T filled into tubes embedded in an agar matrix. Figures 3A/B and 3C (grey bars) show that the temporally unmixed images also provide a clear contrast for Dronpa expressing *E. coli* with respect to blood.

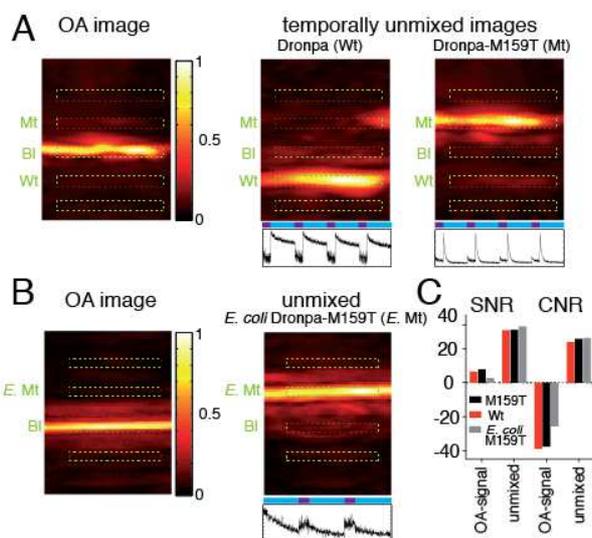

Fig. 3. Temporal unmixing of Dronpa and its variant from background absorbers in optoacoustic images. (**A**) Purified Dronpa-M159T (MT), Dronpa-Wt (Wt), as well as Blood (Bl) were filled in tubings and positioned on the imaging device. Maximum intensity projections of three-dimensional reconstructions are shown for the signals acquired at 488 nm (left panel). The unmixed images of the purified Wt (central panel) and MT (right panel) are displayed together with the time profiles (containing signals recorded at both wavelengths) identified in the unmixing procedure. The photocontrol schedule contained 100 on-switching laser pulses at 420 nm (dark blue) followed by 300 nm readout/off switching (light blue). (**B**) E. coli suspension expressing Dronpa-M159T (E. MT) were, filled in tubings, embedded in an agar phantom and imaged together with blood in an analogous experiment. (**C**) SNR and CNR values for purified proteins (red and black bars) and E. coli suspension (gray bars) were calculated from the ROIs indicated in (A) and (B); the top and bottom ROI are defined as background. The same photocontrol schedule as for the purified protein was used.

In general, temporally unmixed MSOT (tuMSOT) requires a photoabsorbing contrast agent whose photophysical properties can be controlled by external control commands such as photons to program temporal signal profiles of the OA contrast agent that can be differentiated from time-varying background signals. Besides via photoconversion or photoswitching a controlled change in photophysical properties can also be achieved by modulating the excited fluorescent lifetimes of chromophores changing the amount of thermalized energy and thus the optoacoustic signal [19]. The simplest programmed signal trajectory might be obtained from photobleaching or irreversible photoactivation or inactivation of dyes [20]. Instead, the reversibility of photochromic molecules such as RSFPs allows for more complex programmed signal trajectories with many switching cycles. For the majority of currently available RSFPs such as Dronpa, the readout at the excitation wavelength also causes photoswitching of the chromophore thereby limiting possible signal trajectories. Ideally though, the photoswitching is decoupled from the readout of the photophysical properties, as is the case for the RSFP named Dreiklang [21]. To determine the optimal series of commands that generates a desirable time course of the programmed signal, the sensitivity and kinetics of the contrast agent's photoresponse and its photo-fatigue should be taken into account. The programmed signal time course should ideally be uncorrelated from time-varying background signals that in live biological organisms may stem from periodic processes such as blood flow, oxygenation or breathing-induced motion. An additional extension to using the signal from just one or two wavelengths as shown here is to measure more informative features across both the spectral and time dimension, which may enable more efficient unmixing. As for optimizing photochromic molecules for tuMSOT, a large photo-controllable change in one or several of their molar extinction coefficients (ECs) in the NIR spectrum is desirable, as is good resistance to photobleaching. The quantum yield should be minimal or change in an anticorrelated manner with respect to the ECs upon photoswitching. The photocontrol wavelengths should also lie in the NIR, *e.g.* by exploiting 2-photon photoswitching processes as possible in the case of Dronpa [22]. To expand the multiplexing ability of tuMSOT it will also be interesting to explore the range of kinetic profiles and photo-control modes of other RSFPs.

In conclusion, we demonstrated that the photoresponsive photophysical properties of certain RSFPs are suitable to obtain programmed time-varying OA signals. On the example of the RSFP Dronpa and its variant Dronpa-M159T, we have shown that temporal unmixing of their simultaneously switched OA signal trajectories could discriminate well between both proteins despite their similar spectral properties. Both RSFPs could also be unmixed with high CNR from strongly absorbing blood that generated an order of magnitude higher optoacoustic signal. These switchable proteins and their (far) red-shifted variants can thus serve as high-contrast optoacoustic gene reporters for future *in vivo* imaging of *e.g.* zebrafish or mouse models via tuMSOT. Furthermore, the sensitive detection of bacterial suspensions of *E. coli* expressing Dronpa-M159T could become useful for studies of the microbiome, gastrointestinal disease research or for imaging applications such as bacterial tumor tracking [23]. Exploiting the temporal information in addition to common spectral unmixing of optoacoustic signals makes great use of the high frame rate of MSOT systems coupled to fast tunable lasers and adds additional detection and differentiation capabilities to optoacoustic imaging. Exploring and optimizing the large class of synthetic and genetically encoded photochromic molecules, especially in the near-infrared region, in conjunction with optimized photocontrol command schedules could make tuMSOT a new standard for high contrast molecular optoacoustic imaging.

**Acknowledgements:** ACS would like to thank Birte Höcker for constant support and for the opportunity to conduct the protein biochemistry work in her laboratory. We thank Benedikt Küster for very helpful discussions on theoretical aspects of temporal unmixing. G.G.W. acknowledges support by Technische Universität München, the Helmholtz Association of German Research Centers and Helmholtz Alliance ICEMED. We furthermore acknowledge support from the European Research Council under grant agreements ERC-2010-StG-260991 (D.R.) and ERC-2012-StG-311552 (G.G.W).